%% file: EXO-10-006_temp.tex
\pdfoutput=1

\documentclass[11pt,twoside,a4paper,cmspaper,final,collab]{cms-tdr}
\def\svnVersion{57350M}\def\cmsCernNo{2011-062}\def\cmsCernDate{\today}\def\cmsMessage{Submitted to Physics Letters B}

\begin{document}\cmsNoteHeader{EXO-10-006}

\hyphenation{had-ron-i-za-tion}
\hyphenation{cal-or-i-me-ter}
\hyphenation{de-vices}
\RCS$Revision: 56294 $
\RCS$HeadURL: svn+ssh://alverson@svn.cern.ch/reps/tdr2/papers/EXO-10-006/trunk/EXO-10-006.tex $
\RCS$Id: EXO-10-006.tex 56294 2011-05-19 13:08:37Z santanas $

\def\sm{standard model\xspace}
\def\zee{\ensuremath{\text{Z} \to ee}\xspace}
\def\ttbar{\ensuremath{\text{t} \bar{\text{t}}}\xspace}
\def\aeta{\ensuremath{|\eta|}\xspace}
\def\pgev{GeV\xspace}
\def\mgev{GeV\xspace}
\def\st{\ensuremath{S_{\rm T}}\xspace}
\def\pt{\ensuremath{p_{\rm T}}\xspace}
\def\mt{\ensuremath{m_{{\rm T},\text{e}\nu}}\xspace}
\def\mtMu{\ensuremath{m_{T,\mu\nu}}\xspace}
\def\mtLepton{\ensuremath{m_{{\rm T},l\nu}}\xspace}
\def\mee{\ensuremath{m_{\text{ee}}}\xspace}
\def\mej{\ensuremath{m_{\text{ej}}}\xspace}
\def\mtnuj{\ensuremath{m_{{\rm T} \nu \text{j}}}\xspace}
\def\mjj{\ensuremath{m_{\text{jj}}}\xspace}
\def\nmuon{\ensuremath{N_{\mu}}\xspace}
\def\minptmet{$\mbox{min}(p^\text{e}_{\rm T},\mbox{\etmiss})$\xspace}
\def\minptmetMu{$\mbox{min}(p^{\mu}_{\rm T},\mbox{\etmiss})$\xspace}
\def\dphimetele{$|\Delta\phi(\big\slash\hspace{-1.6ex}{E_\text{T}},\text{e})|$\xspace}
\def\dphimetjetone{$|\Delta\phi(\big\slash\hspace{-1.6ex}{E_\text{T}},\mbox{j1})|$\xspace}
\def\dphimetjettwo{$|\Delta\phi(\big\slash\hspace{-1.6ex}{E_\text{T}},\mbox{j2})|$\xspace}
\def\dRejone{$\Delta R(e,\mbox{j1})$\xspace}
\def\dRejtwo{$\Delta R(e,\mbox{j2})$\xspace}
\def\dRejets{$\mbox{min}\Delta R(\text{e},\mbox{jets})$\xspace}
\def \LQ {\mbox{LQ}\xspace}
\def \Lint {$\mathcal{L}_{int}$\xspace}
\def \GeVmass {GeV\xspace}
\def \GeVmom {GeV\xspace}
\def \MLQ {$M_{\text{LQ}}$\xspace}
\def \pp {proton-proton\xspace}
\def \ppbar {proton-antiproton\xspace}
\def \ep {electron-proton\xspace}
\def \sqrts {$\sqrt{s}=7$~TeV\xspace}
\def \etmiss {\ensuremath{E_{\mathrm{T}}\hspace{-1.1em}/\kern0.5em}\xspace}
\def \lumi {36~pb$^{-1}$\xspace}
\def \lumiwitherror {$36.0 \pm 1.5 \pbinv$\xspace} 
\def \lumiPhoton {36~pb$^{-1}$\xspace} 
\def \lumiQCD {6.1~pb$^{-1}$\xspace} 
\def \lumiOpt {30~pb$^{-1}$\xspace} 
\def \LQmassLimitBetaOne {$384$~GeV\xspace} 
\def \LQmassLimitBetaHalf {$320$~GeV\xspace} 
\def \LQmassLimitBetaPointOneComb {255\xspace} 
\def \LQmassLimitBetaHalfComb {340\xspace} 
\def \LQmassLimitBetaOneComb {384\xspace} 
\def \LQmassLimitBetaOneMu {$394$~GeV\xspace} 
\def \LQmassLimitBetaHalfMu {262~GeV\xspace} 
\def \LQmassLimitBetaPointOneCombMu {$<200$\xspace} 
\def \LQmassLimitBetaHalfCombMu {312\xspace} 
\def \LQmassLimitBetaOneCombMu {394\xspace} 
\def \NEventsWithTwoEles {346\xspace}
\def \rescaleFactorW {$1.18 \pm 0.12$\xspace} 
\def \rescaleFactorWMu {$1.20 \pm 0.13$\xspace}
\def \rescaleFactorZ {$1.20 \pm 0.14$\xspace} 
\def \rescaleFactorQCD {$5.4$\xspace} 
\def \ContaminationAtZpeak {$\sim3\%$\xspace} 
\def \ContaminationAtWpeak {$\sim40\%$\xspace} 
\def \ZbkgUncertainty {$12\%$\xspace}  
\def \WbkgUncertainty {$10\%$\xspace}  
\def \TTBARbkgUncertainty {$14\%$\xspace}  
\def \QCDUncertainty {$25\%$\xspace} 
\def \Npileup {2\xspace} 
\def \JSON {Cert\_136033-149442\_7TeV\_Nov4ReReco\_Collisions10\_JSON.txt located at /afs/cern.ch/cms/CAF/CMSCOMM/COMM\_DQM/certification/Collisions10/7TeV/Reprocessing/ and at https://cms-service-dqm.web.cern.ch/cms-service-dqm/CAF/certification/Collisions10/7TeV/Reprocessing/\xspace}
\def \MinLQPDFerr{$\pm8$\%\xspace}
\def \MaxLQPDFerr{$\pm22$\%\xspace}
\def \LQmassMinPDFerr{$200$\xspace}
\def \LQmassMaxPDFerr{$500$\xspace}
\def \MLQforSystematics{$300$\xspace}
\def \MLQforSelection{$300$\xspace}
\def \MLQforStPlot{$200$\xspace}
\def \LQmassesForPlots{250, 300, and 340\xspace}
\def \singleEleSelEffForMLQForSelection{$\sim 80\%$\xspace}
\def \singleEleSelEffForMLQForSelectionMin{$\sim 76\%$\xspace}
\def \singleEleSelEffForMLQForSelectionMax{$\sim 83\%$\xspace}
\def \singleMuonSelEffForMLQForSelection{$83\%$\xspace}
\def \EESEB{1\%\xspace}
\def \EESEE{3\%\xspace}
\def \JES{5\%\xspace}
\def \ptcutForMetscale{30\xspace}
\def \NLQsamples{10\xspace} 
\def \minLQmassForSamples{200\xspace} 
\def \maxLQmassForSamples{500\xspace} 

\def \eejj{$\text{eejj}$\xspace}
\def \mumujj{$\mu\mu \text{jj}$\xspace}
\def \enujj{$\text{e}\nu \text{jj}$\xspace}
\def \munujj{$\mu\nu \text{jj}$\xspace}
\def \nunujj{$\nu\nu \text{jj}$\xspace}
\def \zjets{\ensuremath{\text{Z}/\gamma}+jets\xspace}
\def \wjets{\ensuremath{\text{W}}+jets\xspace}
\def \MLQforMejPlotInAppendix {$300$\xspace} 
\def \PrelimStCut {250~GeV\xspace}  
\def \PrelimMetCutMu {45~GeV\xspace}
\def \PrelimPtCutMu  {35~GeV\xspace}
\def \LQMassesAnalyzed{from 200 to 500 GeV\xspace}
\def \MtFinalCut{$ > 125$\xspace}
\def \PDFUncertaintyOnLQCrossSection{from 8 to 22\% for LQ masses from 200 to 500 GeV\xspace}
\def \theoreticalUncertaintyOnLQCrossSection{from 13 to 15\% for all considered LQ masses\xspace}
\def \LQmassLimitBetaHalfWithTheoryError {$310$~GeV\xspace} 
\def \LQmassLimitBetaHalfWithTheoryErrorMu {$252$~GeV\xspace} 

\def \EMIso{$\text{EM}_\text{Iso}$\xspace}
\def \HADIsoOne{$\text{HAD}^{\text{layer1}}_\text{Iso}$\xspace}
\def \HADIsoTwo{$\text{HAD}^{\text{layer2}}_\text{Iso}$\xspace}

\cmsNoteHeader{EXO-10-006} 
\title{Search for First Generation Scalar Leptoquarks \\ in the $\text{e}\nu \text{jj}$ Channel in pp Collisions at $\sqrt{s} =$ 7 TeV} 

\address[Maryland]{U. Maryland}
\author[cern]{The CMS Collaboration}

\date{\today}

\abstract{
A search for pair-production of first generation scalar leptoquarks
is performed in the final state containing an electron, a neutrino, and
at least two jets using proton-proton collision data at $\sqrt{s}=7$~TeV. The data were collected
by the CMS detector at the
LHC, corresponding to an
integrated luminosity of 36~pb$^{-1}$. The number of observed events is in good
agreement with the predictions for standard model processes.
Prior CMS results in the dielectron channel are combined with this
electron+neutrino search.  A 95\% confidence level combined lower limit is set on
the mass of a first generation scalar leptoquark at 340~GeV
for $\beta=0.5$, where $\beta$ is the
branching fraction of the leptoquark to an electron and a quark.~These
results represent the most stringent direct limits to
date for values of $\beta$ greater than 0.05.
}

\hypersetup{%
pdfauthor={CMS Collaboration},%
pdftitle={Search for First Generation Scalar Leptoquarks in the evjj channel in pp collisions at sqrt s = 7 TeV},%
pdfsubject={CMS},%
pdfkeywords={LHC, CMS, physics, exotica, leptoquarks}}

\maketitle 

\input{introduction.tex}
\input{detector.tex}

\input{samples.tex}

\input{selection.tex}

\input{backgrounds.tex}

\input{systematics.tex}

\input{results.tex}

\input{conclusions.tex}

\bibliography{auto_generated}   

\cleardoublepage\appendix\section{The CMS Collaboration \label{app:collab}}\begin{sloppypar}\hyphenpenalty=5000\widowpenalty=500\clubpenalty=5000\input{EXO-10-006-authorlist.tex}\end{sloppypar}
\end{document}

%% file: introduction.tex
\section{Introduction}

The \sm (SM) of particle physics has an intriguing 
but 
ad hoc symmetry between quarks and leptons.  
In some 
theories 
beyond the SM, 
such as SU(5) grand unification~\cite{PhysRevLett.32.438}, 
Pati--Salam SU(4)~\cite{Pati:1974yy},
composite models~\cite{Schrempp:1984nj}, 
technicolor~\cite{Dimopoulos:1979es,Dimopoulos:1979sp,Farhi:1980xs},
and super\-string-inspired $E_6$ models~\cite{Hewett:1988xc},  
the existence of a new symmetry 
relates the quarks and leptons 
in a fundamental way. These models predict
the existence of new bosons, called leptoquarks.
The leptoquark (LQ) is coloured, has fractional electric charge, 
and couples to a lepton and a quark with coupling strength
$\lambda$.  
The leptoquark decays to a charged lepton and a quark, with unknown branching fraction
$\beta$, or a neutrino and a quark, with branching fraction
$(1-\beta)$.
A review of LQ phenomenology and searches can be found in Refs.~\cite{Acosta:1999ws,Hewett:1997ce}.
Constraints from experiments sensitive to flavour-changing
neutral currents, lepton-family-number violation,
and other rare processes~\cite{Buchmuller:1986iq}
favour LQs that couple to quarks and leptons within the same SM generation,
for LQ masses accessible to current colliders.

The first generation scalar LQs studied in this paper 
have spin 0 and couple only
to electron or electron neutrino and up or down quark.
Measurements at the HERA \ep collider constrain the coupling 
$\lambda$ to be less than the electromagnetic coupling
for LQ mass, $M_{\rm LQ}$, less than 300~\GeVmass~\cite{Aktas:2005pr,Chekanov:2003af}.
Prior to the results of the Large Hadron Collider (LHC) experiments,
direct limits on the mass of the first generation scalar LQ have also 
been set by the Tevatron~\cite{Abazov:2009gf,Acosta:2005ge} and 
LEP~\cite{Abbiendi:2003iv,tagkey19931,Abreu1993620,tagkey1992253} 
experiments, for a broad range of the coupling $\lambda$.
The Compact Muon Solenoid (CMS) experiment 
published a stricter lower limit of \LQmassLimitBetaOneComb GeV~\cite{PhysRevLett.106.201802}
on the mass of first generation scalar LQs for $\beta=1$ in the 
dielectron-plus-dijet (\eejj) channel, 
using a sample collected in \pp collisions at \sqrts and corresponding 
to an integrated luminosity of approximately 33~pb$^{-1}$.
Recently, the ATLAS experiment at the LHC has also obtained an exclusion 
on the mass of first generation scalar LQs~\cite{Collaboration:2011uv}.

This paper presents the results of a search for pair-production of 
first generation scalar LQs using events containing an electron, 
missing transverse energy, and at least two jets (\enujj) 
using \pp collision data at $\sqrt{s} = 7$ TeV.
In \pp~collisions at the LHC, LQs are predominantly 
pair-produced via gluon-gluon fusion with a
cross section that depends on the strong coupling
constant $\alpha_{\rm s}$ but is nearly independent of $\lambda$. 
This cross section depends on the spin and the mass
of the LQ and has been calculated at the next-to-leading-order (NLO)
in QCD~\cite{PhysRevD.71.057503}.
LQs could also be produced singly with a cross section that is dependent
on $\lambda$. The results of this study are based on the assumption 
that $\lambda$ is sufficiently small that single-LQ production can be neglected.
The data were collected in 2010 by the CMS detector at the CERN LHC 
and correspond to an integrated luminosity 
of \lumi.
The \eejj and \enujj combined results are also presented.

%% file: detector.tex
\section{The CMS Detector}
The CMS experiment uses a right-handed coordinate system, with the origin at the nominal 
interaction point, the $x$-axis pointing to the centre of the LHC, the $y$-axis pointing 
up (perpendicular to the LHC plane), and the $z$-axis along the anticlockwise-beam direction. 
The polar angle $\theta$ is measured from the positive $z$-axis and the azimuthal angle 
$\phi$ is measured in the $xy$ plane.
Pseudorapidity is defined as $\eta = -\ln[\tan(\theta/2)]$.
The central feature of the CMS apparatus is a
superconducting solenoid, of 6~m internal diameter, providing a field
of 3.8~T. Within the field volume are a silicon pixel and strip
tracker, a crystal electromagnetic calorimeter (ECAL), and a
brass/scintillator hadron calorimeter (HCAL). Muons are measured in
gas-ionization detectors embedded in the steel return yoke. In
addition to the barrel and endcap detectors, CMS has extensive forward
calorimetry.
The ECAL has an energy resolution of better than 0.5\% for 
unconverted photons with transverse energies above 100~\GeVmom, 
and 3\% or better for electrons of energies relevant to this analysis.
In the region $\vert \eta \vert< 1.74$, the HCAL cells have 
granularity $\Delta \eta \times \Delta \phi = 0.087\times0.087$
(where $\phi$ is measured in radians).
In the $(\eta,\phi)$ plane, and for $\vert \eta \vert< 1.48$, the 
HCAL cells map on to $5 \times 5$ ECAL crystals arrays to form 
calorimeter towers projecting radially outwards from close to the 
nominal interaction point. At larger values of $\vert \eta \vert$, 
the size of the towers increases and the matching ECAL arrays contain 
fewer crystals.
The muons are measured in the pseudorapidity window $|\eta|< 2.4$,
with detection planes made of three technologies: drift tubes, cathode
strip chambers, and resistive plate chambers. Matching the muons to
the tracks measured in the silicon tracker results in a transverse
momentum ($p_{\rm T}$) resolution between 1 and 5\%, for $p_{\rm T}$ values up to
1~TeV.
The inner tracker measures charged particles within $|\eta| < 2.5$. 
It consists of 1440 silicon pixel and 15\,148
silicon strip detector modules and 
provides an impact parameter resolution of $\sim$\,15~$\mu$m and a
$p_{\rm T}$ resolution of about 1.5\% for
100~GeV particles.
Events must pass a first-level trigger made of a system of
fast electronics and a high-level trigger that consists
of a farm of commercial CPUs running a version of the offline
reconstruction software optimized for timing considerations.
A detailed description of the CMS detector can be found 
elsewhere~\cite{:2008zzk}.

%% file: samples.tex
\section{Reconstruction of Electrons, Muons, Jets, and $E_{\mathrm{T}}\hspace{-0.9em}/\kern0.5em$}

Events used in the \enujj analysis are 
collected by single-electron
triggers without isolation requirements and 
with \pt thresholds dependent upon the running period, 
because of the evolving beam conditions during 2010.
The bulk of the data were collected with a trigger requiring 
an electron with $\pt > 22$~GeV.
Events are required to contain at least one primary
vertex with reconstructed $z$ position within 24 cm, and $xy$ position 
within 2 cm of the nominal center of the detector.

Electron candidates 
are required to have an electromagnetic cluster in ECAL
that is spatially matched to a reconstructed track
in the central tracking system in both $\eta$ and $\phi$, and
to have a shower shape consistent with that of an electromagnetic shower. 
The ratio $H/E$, where $E$ is the energy of the ECAL cluster and 
$H$ is the energy in the HCAL cells situated behind it, within 
a cone of radius $\Delta R = \sqrt{ (\Delta \phi)^2 + (\Delta \eta)^2} = 0.15$
centred on the electron, is required to be less than 5\%. 
Electron candidates are further required 
to be isolated from additional energy deposits 
in the calorimeter and from additional reconstructed tracks (beyond the matched track) 
in the central tracking system. 
The sum of the \pt of the tracks in an hollow 
cone of internal (external) radius $\Delta R = 0.04$ (0.3) is required
to be less than 7.5 (15)~\GeVmom for electron candidates reconstructed within 
the barrel (endcap) acceptance. The ECAL isolation 
variable, \EMIso, is defined as the sum of the transverse 
energy in ECAL cells within a cone of radius $\Delta R = 0.3$. To remove the contribution 
from the electron itself, the sum is performed excluding ECAL 
energy deposits in an inner cone of radius 3 crystals and in 
a strip, with a total width of 3 crystals in $\eta$ 
and $2 \times 0.3$ radians in $\phi$, both centred on the electron position.
The longitudinal segmentation of the HCAL calorimeter 
is exploited in the isolation. The HCAL isolation 
variables, \HADIsoOne and \HADIsoTwo, are defined as the sum of 
transverse energy deposits in an hollow cone of internal (external) 
radius $\Delta R = 0.15$ (0.3), where the sum is performed
over the first and second readout layers of the HCAL calorimeter, respectively.
In the barrel, where only one HCAL layer is present, electron candidates are 
required to have \EMIso~$+$~\HADIsoOne less than $0.03 \, p_{{\rm T},\text{e}} + 2$~\GeVmass.
In the endcaps, candidates with $p_{{\rm T},\text{e}}$ below (above) 50~\GeVmom 
are required to have \EMIso~$+$~\HADIsoOne less than 2.5~\GeVmass 
($0.03 \, [ p_{{\rm T},\text{e}} - 50 ] + 2.5$~\GeVmass);
the isolation variable \HADIsoTwo is also required to be less than $0.5$~\GeVmass, 
independent of the electron \pt. 
Electrons reconstructed near the crack between ECAL barrel 
and endcap ($1.44<|\eta|<1.56$) are not considered.
More information about electron identification at CMS
during this running period can be found in Ref.~\cite{EGM-10-004}.

Jets
are reconstructed by the anti-k$_\mathrm{T}$ algorithm~\cite{antikt} 
from a list of particles obtained using particle-flow 
methods and a radius parameter $R$ = 0.5.
The particle-flow 
algorithm~\cite{PFPAS2009} 
reconstructs a complete, unique list of particles
in each event using an optimized combination of information
from all CMS subdetector systems.
Particles that are reconstructed and identified include muons,
electrons (with associated bremsstrahlung photons), photons (unconverted and
converted), and charged/neutral hadrons.
The jet energy corrections are derived using Monte Carlo (MC) simulation
and {\it in situ} measurements using dijet and photon+jet events~\cite{PAS-JME-10-010}.  

The transverse momentum of the neutrino is estimated from the 
missing transverse energy \etmiss, 
which is the magnitude of the negative vector sum
of all particle-flow objects' transverse momenta. 
More information about \etmiss performance during this running period 
can be found in Refs.~\cite{JME-10-004,JME-10-005,PFPAS2010}.

Muon candidates 
are reconstructed as tracks in the muon system
that are spatially matched to a reconstructed track in the inner tracking system.
To ensure a precise measurement of the impact parameter, only 
muons with tracks containing at least 11 hits 
in the silicon tracker are considered. 
To reject cosmic muons, the transverse impact parameter 
with respect to the beam axis is required to be less than 2 mm.
The relative isolation parameter is defined  
as the scalar sum of the \pt of all tracks in the tracker and the 
transverse energies of hits in the ECAL and HCAL in a cone of 
$\Delta R = 0.3$ around the muon track, excluding the contribution from the 
muon itself, divided by the muon \pt. 
Muons are required to have a relative isolation value less than 5\%. 
A veto on the presence of isolated muons in the final state 
is used to reject \ttbar background events, as described later.

%% file: selection.tex
\section{Event Samples and Selection}

\subsection{MC Samples} \label{sec:MCsamples}

The dominant sources of \enujj events from production of \sm particles are
pair-production of top quarks and associated production of a W boson with jets.
There is also a small contribution from
multijet events with a jet misidentified
as an electron and spurious missing transverse energy
due to mismeasurement of jets, associated production of Z boson with jets,
in addition to single top, diboson, b+jets, and $\gamma$+jets production.

To compare collision data to MC, the response of the detector was simulated using
{\sc Geant4}~\cite{GEANT4}. The detector geometry description included
realistic subsystem conditions such as defunct and noisy channels.
The selection procedure as well as the electron, muon, jet, and \etmiss
reconstructions described for the data are also applied to the MC simulation samples.
For the generation of all the MC samples, the CTEQ6L1~\cite{1126-6708-2002-07-012}
parton distribution functions (PDFs) were used.
The \wjets and \zjets events were generated using {\sc alpgen}~\cite{Mangano:2002ea}.
The \ttbar, single-top, b+jets, and $\gamma$+jets events were generated
using {\sc MadGraph}~\cite{madgraph,Alwall:2007st}.
The diboson (WW, ZZ, WZ) events were generated
using {\sc pythia}~\cite{pythia}, version 6.422,
tune D6T~\cite{Field:2008zz,Field:2010su}.
For the {\sc MadGraph} and {\sc alpgen} samples, parton showering and
hadronization were performed with {\sc pythia}.
The QCD multijet background is estimated from data, as described later.
Signal samples for LQ masses ($M_{\text{LQ}}$) \LQMassesAnalyzed
were generated with {\sc pythia}.
The product of single-electron efficiency and acceptance, requiring
a minimum electron \pt of 35 \GeVmom,
varies from \singleEleSelEffForMLQForSelectionMin to \singleEleSelEffForMLQForSelectionMax
for LQ masses \LQMassesAnalyzed.

\subsection{Preselection}

Samples enriched in the aforementioned SM
processes are selected to verify the background estimate.
The \enujj preselection requires exactly one electron
with \pt$> 35$ \pgev and $|\eta|<2.2$, at
least two jets with \pt $>$ 30 \pgev and $|\eta|<3.0$,
and \etmiss $> 45$ \pgev.
The electron is also required to be separated from both
the two leading jets by a distance $\Delta R > 0.7$.
In addition, to reduce the \ttbar background,
events with at least one isolated muon with \pt$>10$~\pgev are rejected.
To reduce the contribution from multijet events
and, in general, events with misidentified \etmiss due to jet mismeasurement,
the opening angle in $\phi$ between the \etmiss vector and the electron
($\Delta \phi_{\text{e}\nu}$), and between the \etmiss vector and the
leading (in \pt) jet are required to be greater than 0.8 and 0.5 radians, respectively.
In addition, a preselection requirement
$\st >$~\PrelimStCut is applied, where \st
is defined as the scalar sum of the \pt of the electron, the \pt of the two
leading jets, and the \etmiss. This variable has a large signal-to-background
discrimination power since the LQ decay products usually have large \pt.

\begin{figure*}[htbp]
  \begin{center}
    \begin{tabular}{cc}
      \includegraphics[width=0.49\textwidth]{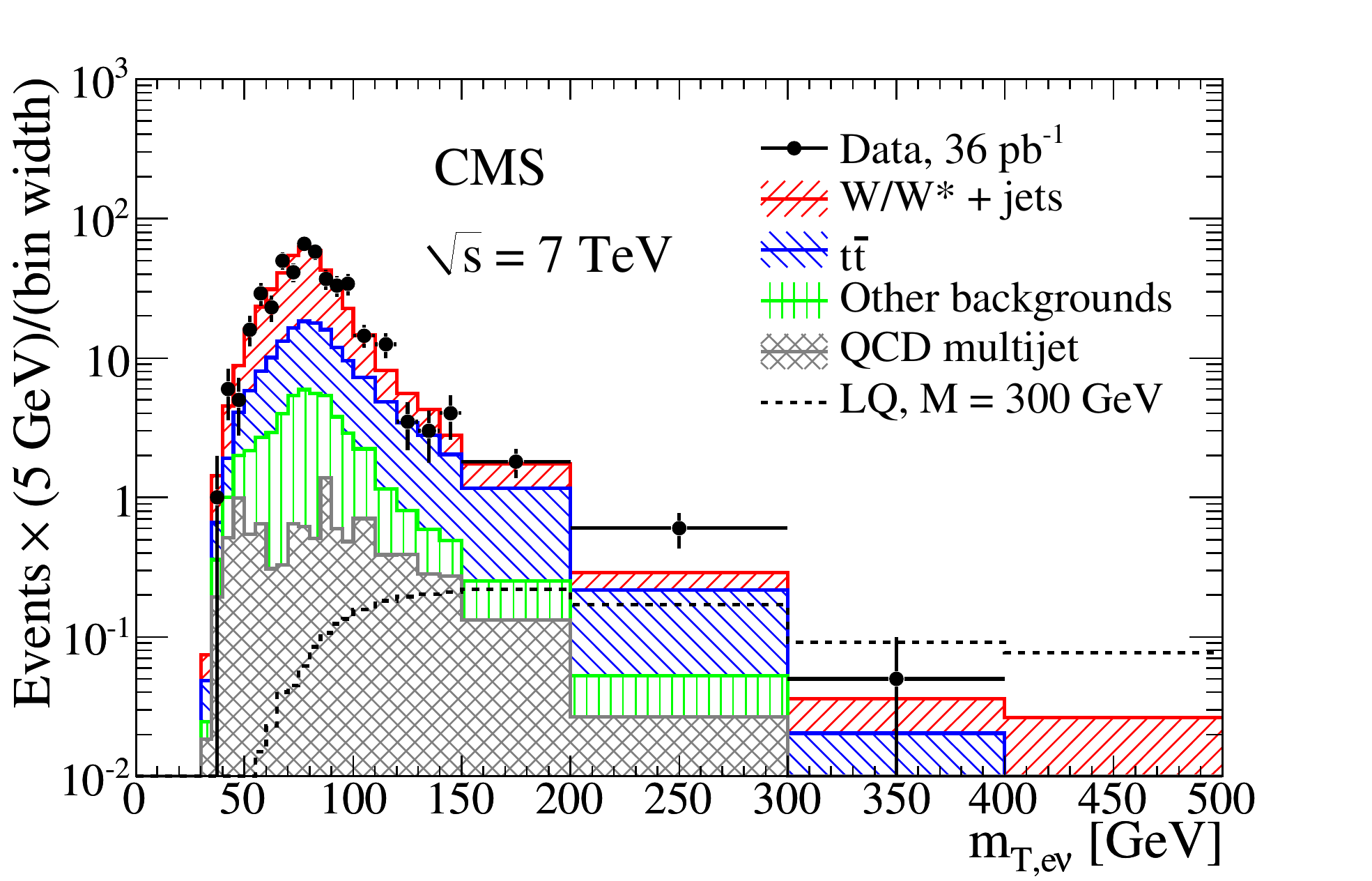} &
      \includegraphics[width=0.49\textwidth]{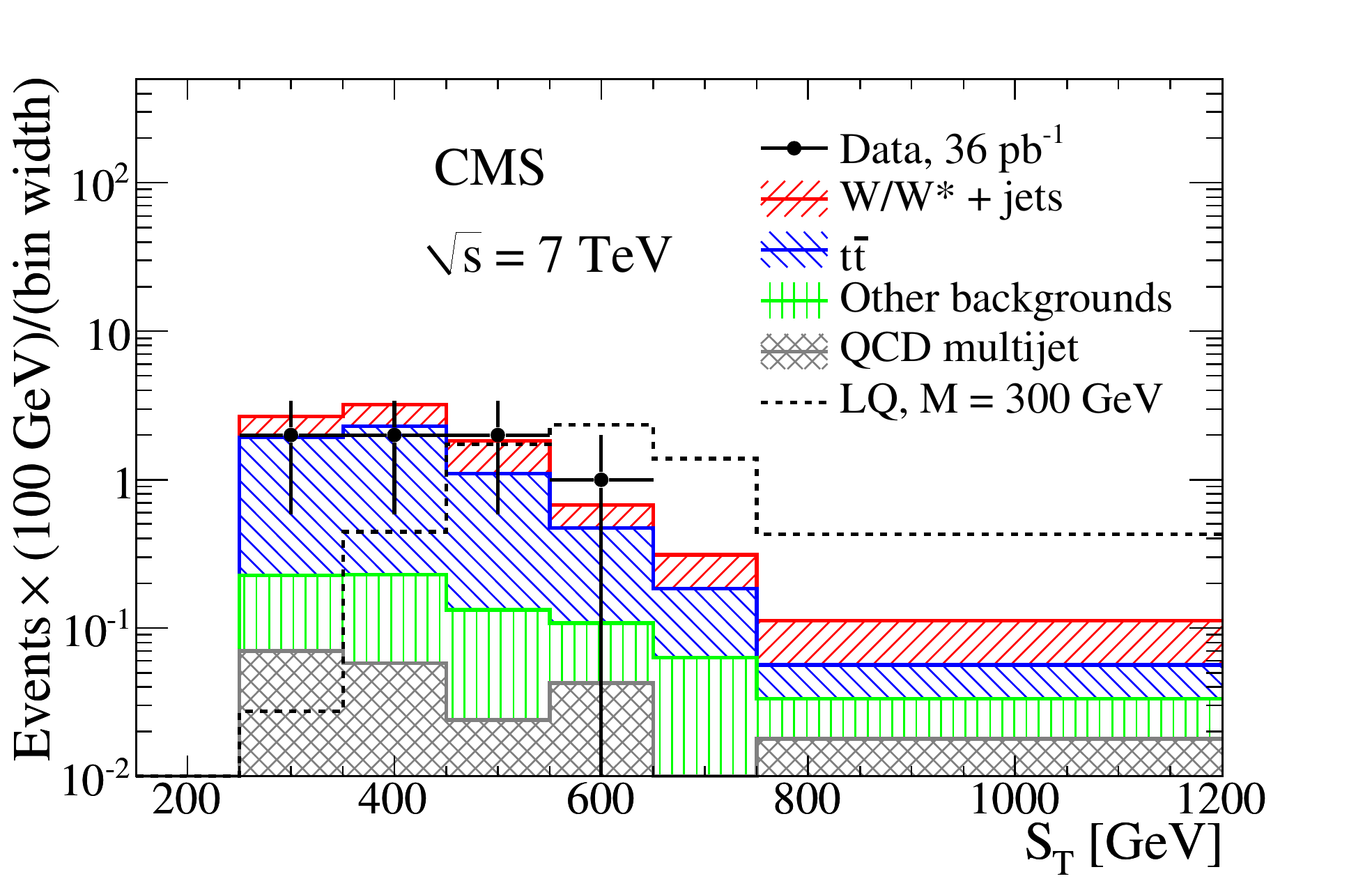} \\
          \end{tabular}
	  \caption{(Left) The \mt distribution for events passing
		   the \enujj preselection requirements.
		   (Right) The \st distribution for events passing the final
		   \enujj event selection except the optimized \st cut itself.
		   The MC distributions for the signal
		   ($M_{\text{LQ}}=$~\MLQforSelection \GeVmass, $\beta=0.5$)
		   and the contributing backgrounds listed in Section~\ref{sec:MCsamples} are shown.
		   }
    \label{fig:AfterPresel}
  \end{center}
\end{figure*}

A sufficient number of data events survive the preselection
to allow a comparison with the background predictions.
A good agreement is observed between data and
background predictions in the shape of all kinematic
distributions of the electron, \etmiss, and jets.
Figure~\ref{fig:AfterPresel} (left) shows the distribution of
the transverse mass of the electron and the neutrino,
defined as
$\mbox{\mt} = \sqrt{ 2 \, p_{{\rm T},\text{e}} \, \mbox{\etmiss} \, (1-\cos{ (\Delta \phi_{\text{e}\nu}) })}$,
after the preselection.
The normalization of the various background sources
is discussed in Section~\ref{sec:background}.

\subsection{Final Selection}

To further reduce backgrounds, the selection criteria are optimized
by minimizing the expected upper limit on the leptoquark cross section in the
absence of signal using a Bayesian approach~\cite{Bertram:2000br}
that is well suited for counting experiments in the Poisson regime.
The final selection requires electron \pt$>85$~\GeVmom,
\etmiss$>85$~\GeVmom, and \mt\MtFinalCut~\GeVmass.
The optimum value of the requirement on \st
was found to vary with the assumed LQ mass.
An alternative discovery optimization that maximizes
the significance estimator $\text{S}/\sqrt{\text{S}+\text{B}+{\sigma_{\text{B}}}^2}$,
where S (B) is the number of signal (background)
events passing the full selection and $\sigma_{\text{B}}$ is
the systematic uncertainty on the background, gives similar results.

Table~\ref{tab:cutFlow}
shows the number of events
surviving the different stages of the \enujj event selection,
for \MLQforSelection \GeVmass mass LQ signal, background, and data samples.
Figure \ref{fig:AfterPresel} (right) shows
the distribution of the \st variable after the full selection
except the optimized \st cut itself.
Table~\ref{tab:NeventFinal}
shows the number of surviving events for MC signal, background,
and data samples after applying the full selection optimized
for different LQ mass hypotheses.
The signal selection efficiencies reported in Table~\ref{tab:cutFlow} and Table~\ref{tab:NeventFinal}
include the kinematic acceptance and are estimated from MC studies.
The systematic uncertainties on the LQ selection efficiency are discussed in Section~\ref{sec:systematics}.

\begin{table*}[htbp]
\caption{Number of \enujj events for the first generation LQ signal
(\MLQforSelection \GeVmass mass, $\beta=0.5$), background, and data samples after
each step of the event selection optimized for \MLQforSelection \GeVmass mass LQ signal.
All uncertainties are statistical only.
The product of signal acceptance and efficiency is also
reported (the statistical uncertainty is less than 1\%).
}
\label{tab:cutFlow}
\begin{center} \scriptsize 
\begin{tabular}{|l|cc|ccccc|c|}
\hline
Cut  & \multicolumn{2}{c|}{MC LQ300 Sample} & \multicolumn{5}{c|}{MC and QCD Background Samples} & Events \\
     & Selected & Acceptance  & \multicolumn{5}{c|}{Selected Events in}  & in\\
     & Events   & $\times$ Efficiency & $\text{t}\bar{\text{t}}$ + jets   & W + jets   & Other Bkgs   & QCD  & All Bkgs & Data\\
\hline \hline
 \enujj preselection & 11.52$\pm$0.03 & 0.529 & 132.9$\pm$0.7 & 306$\pm$3 & 44.6$\pm$0.6 & 13.7$\pm$0.4 & 497$\pm$4 & 505 \\
 \mt$>125$ \GeVmass & 10.01$\pm$0.03 & 0.459 & 22.7$\pm$0.3 & 14.2$\pm$0.8 & 3.3$\pm$0.2 & 3.5$\pm$0.2 & 43.6$\pm$0.9 & 46 \\
 \minptmet$>85$ \GeVmom & 7.89$\pm$0.03 & 0.362 & 5.3$\pm$0.2 & 3.0$\pm$0.4 & 0.63$\pm$0.06 & 0.27$\pm$0.05 & 9.2$\pm$0.4 & 7 \\
 \st$>490$ \GeVmom & 6.89$\pm$0.03 & 0.317 & 1.09$\pm$0.07 & 1.0$\pm$0.2 & 0.27$\pm$0.05 & 0.14$\pm$0.04 & 2.5$\pm$0.2 & 2 \\
\hline
\end{tabular}
\end{center}
\end{table*}

\begin{table*}[htbp]
\caption{Number of \enujj events for the first generation LQ signal ($\beta=0.5$), background,
and data samples after the full analysis selection.
The optimum value of the requirement on \st
is reported in the first column for each LQ mass.
All uncertainties are statistical only.
The product of signal acceptance and efficiency is also
reported for different LQ masses (the statistical uncertainty is less than 1\%).
}
\label{tab:NeventFinal}
\begin{center} \scriptsize 
\begin{tabular}{|c|cc|ccccc|c|}
\hline
$M_{\text{LQ}}$  & \multicolumn{2}{c|}{MC Signal Samples} & \multicolumn{5}{c|}{MC and QCD Background Samples} & Events \\
(\st~cut) & Selected & Acceptance  & \multicolumn{5}{c|}{Selected Events in}  & in\\
{[GeV]}     & Events   & $\times$ Efficiency & $\text{t}\bar{\text{t}}$ + jets   & W + jets   & Other Bkgs   & QCD  & All Bkgs & Data\\
\hline \hline
 200 (\st$>350$) & 34.5$\pm$0.2    & 0.161 & 3.6$\pm$0.1   & 2.2$\pm$0.3   & 0.48$\pm$0.06 & 0.20$\pm$0.04 & 6.5$\pm$0.3   & 5 \\
 250 (\st$>410$) & 15.9$\pm$0.1  & 0.255 & 2.24$\pm$0.09 & 1.7$\pm$0.3   & 0.35$\pm$0.05 & 0.18$\pm$0.05 & 4.4$\pm$0.3   & 3 \\
 280 (\st$>460$) & 9.54$\pm$0.05   & 0.291 & 1.43$\pm$0.08 & 1.2$\pm$0.2   & 0.29$\pm$0.05 & 0.14$\pm$0.04 & 3.1$\pm$0.2   & 3 \\
 300 (\st$>490$) & 6.89$\pm$0.03   & 0.317 & 1.09$\pm$0.07 & 1.0$\pm$0.2   & 0.27$\pm$0.05 & 0.14$\pm$0.04 & 2.5$\pm$0.2   & 2 \\
 320 (\st$>520$) & 5.03$\pm$0.02   & 0.339 & 0.75$\pm$0.05 & 0.8$\pm$0.2   & 0.22$\pm$0.05 & 0.13$\pm$0.04 & 1.9$\pm$0.2   & 2 \\
 340 (\st$>540$) & 3.73$\pm$0.02   & 0.364 & 0.65$\pm$0.05 & 0.7$\pm$0.2   & 0.20$\pm$0.05 & 0.12$\pm$0.04 & 1.6$\pm$0.2   & 2 \\
 370 (\st$>570$) & 2.40$\pm$0.01 & 0.396 & 0.50$\pm$0.04 & 0.6$\pm$0.1   & 0.18$\pm$0.04 & 0.08$\pm$0.03 & 1.3$\pm$0.2   & 1 \\
 400 (\st$>600$) & 1.57$\pm$0.01 & 0.426 & 0.34$\pm$0.04 & 0.5$\pm$0.1   & 0.17$\pm$0.04 & 0.08$\pm$0.03 & 1.1$\pm$0.1   & 1 \\
 450 (\st$>640$) & 0.797$\pm$0.003 & 0.467 & 0.26$\pm$0.03 & 0.4$\pm$0.1   & 0.13$\pm$0.04 & 0.08$\pm$0.04 & 0.9$\pm$0.1   & 0 \\
 500 (\st$>670$) & 0.417$\pm$0.001 & 0.500 & 0.18$\pm$0.03 & 0.4$\pm$0.1   & 0.12$\pm$0.04 & 0.08$\pm$0.04 & 0.8$\pm$0.1   & 0 \\
\hline
\end{tabular}
\end{center}
\end{table*}

%% file: backgrounds.tex
\section{Backgrounds} \label{sec:background}

The \ttbar background is estimated from MC 
assuming an uncertainty on the inclusive 
\ttbar production cross section of \TTBARbkgUncertainty, 
taken from the CMS measurement~\cite{TOP-11-002}. 
Since the latter is consistent with next-to-next-to-leading-logarithm 
(NNLL) predictions, no rescaling of the \ttbar MC sample is applied. 
The small contribution from Z+jets, single top, diboson,
b+jets, and $\gamma$+jets is estimated via MC. 

The QCD multijet background is determined from data. 
The probability of an isolated electromagnetic cluster being 
reconstructed as an electron is measured in a 
sample with at least two jets and small \etmiss.
For $|\eta|<1.44$, this probability is found to 
be $\sim 5 \times 10^{-3}$, independent of the 
transverse energy deposit of the cluster.
For $1.56<|\eta|<2.2$, this 
probability grows linearly as a function of cluster \pt, varying 
between $\sim 2 \times 10^{-2}$ and $\sim 4 \times 10^{-2}$ for 
cluster \pt between 50 and 200 GeV.
This probability is applied to a sample with one cluster, large \etmiss,
and two or more jets to predict the QCD multijet contribution to the final 
selection sample.
The systematic uncertainty is determined to be \QCDUncertainty, by using 
probabilities derived in samples with different number of jets (more than 1 or 3)
and by calculating the maximum relative variation in the 
number of background events predicted at the preselection level.
This background accounts for $\sim 5\%$ of the total background 
for the LQ masses of interest.

The \wjets background dominates the \enujj preselection sample. 
At this level of the selection, 
the ratio $R_{{\rm W}}=(N_{{\rm data}}-N_{{\rm OB}})/N_{{\rm W}} $ 
is calculated, where $N_{{\rm data}}$, $N_{{\rm W}}$, and $N_{{\rm OB}}$ are the numbers of events 
in data, \wjets, and other MC backgrounds with $50<\mbox{\mt}<110$~\GeVmass.
The value of $R_{{\rm W}}$ 
is \rescaleFactorW; this rescaling factor is used to normalize the \wjets MC sample.
The relative uncertainty on this normalization factor, which depends both on the statistical uncertainty 
on the data and on the systematic uncertainties on the other backgrounds 
contaminating the preselection sample, is used as the uncertainty 
on the MC estimate of the \wjets background. 
In addition, an uncertainty on the modeling of the 
shape of this background is determined using {\sc MadGraph} samples 
with renormalization and factorization scales and jet-matching thresholds 
at the generator level varied by a factor of two in each direction.
The number of \wjets events surviving the preselection criteria and the final \mt cut
is compared among the aforementioned MC samples, and the largest deviation 
from the default value is used to assess a 44\% systematic uncertainty.

%% file: systematics.tex
\section{Systematic Uncertainties} \label{sec:systematics}

The impact of the systematic uncertainties on
the numbers of signal and background events
is summarized in Table~\ref{tab:SysUncertainties}.
The uncertainties on the \ttbar and \wjets normalization,
and the \wjets background shape, are discussed in Section~\ref{sec:background}.
For the energy scales of electrons and jets,
the event selection is repeated
with the jet and electron energies
rescaled by
by a factor of $1 \pm \delta$,
where $\delta$ is the relative uncertainty on their energy scales.
The uncertainty on the \etmiss scale is
primarily affected by the uncertainty on the jet energy scale.
The event-by-event variation in the \etmiss and jet measurements, due to the relative changes in the energies
of the reconstructed jets, is used to determine the quoted energy scale uncertainty of jets and \etmiss.
The statistical uncertainty on the number of \enujj MC events, after
the full selection, is summarized in Table~\ref{tab:NeventFinal}
for signal and background samples.
The systematic uncertainty on trigger, reconstruction, identification,
and isolation efficiency for electrons is assessed
using $\text{Z} \rightarrow \text{ee}$ events from data, using methods similar
to those discussed in Ref.~\cite{Chatrchyan:2011wq}.
The uncertainty on
the integrated luminosity of the data sample
is 4\%~\cite{LumiNote4Percent}.
The effect of the PDF uncertainties on the signal acceptance
is estimated using an event reweighting technique that uses the LHAPDF
package~\cite{Bourilkov:2006cj} and it is found to be negligible (less than 1\%).
For the dominant \ttbar and \wjets backgrounds
the uncertainties due to the PDF choice, electron efficiencies,
and integrated luminosity are not considered, as those effects
are included in the normalization uncertainty.

\begin{table*}[htbp]
\caption{Summary of the systematic uncertainties on the numbers of signal and
background events for a LQ with mass \MLQforSystematics \GeVmass.}
\label{tab:SysUncertainties}
\begin{center}
\begin{tabular}{|l|c|c|c|}
\hline
Source      & Systematic            & Effect on     & Effect on             \\
            & Uncertainty $[\%]$    & Signal $[\%]$ & Background $[\%]$ \\
\hline\hline
\ttbar (\wjets) Normalization   & 14 (10)     &   -   &   10   \\
\wjets Background Shape    &   44    &   -   &   17   \\
Jet/\etmiss Energy Scale   &   5   &   5   &   7   \\
Electron Energy Scale Barrel (Endcap)   &   1 (3)   &   1   & 3   \\
MC Statistics   &   [Table~\ref{tab:NeventFinal}]  &   0.4   &   9   \\
Electron Trigger/Reco/ID/Isolation   &   6   &   6   &   -   \\
Integrated Luminosity          &   4   &   4   &   -   \\
\hline\hline
\multicolumn{2}{|l|}{Total}   &   9   &   23   \\
\hline
\end{tabular}
\end{center}
\end{table*}

%% file: results.tex
\section{Results}

The number of observed events in data passing the full selection criteria
is consistent with the prediction from SM processes, as reported in the last two columns of
Table~\ref{tab:NeventFinal}.
An upper limit on the LQ cross section in the absence of the leptoquark signal
is therefore set using a Bayesian approach~\cite{Bertram:2000br}
featuring a flat signal prior, Poisson statistics,
and log-normal priors to integrate over
the systematic uncertainties marginalized as nuisance parameters.
The systematic uncertainties for the background are dominated by
the \ttbar and \wjets normalization uncertainty and the uncertainty
on the \wjets background shape.
Systematic uncertainties on the signal efficiency are dominated
by the uncertainty on the electron selection efficiencies and the jet energy scale.

\begin{figure*}[htbp]
  \centering
  \includegraphics[width=0.49\textwidth]{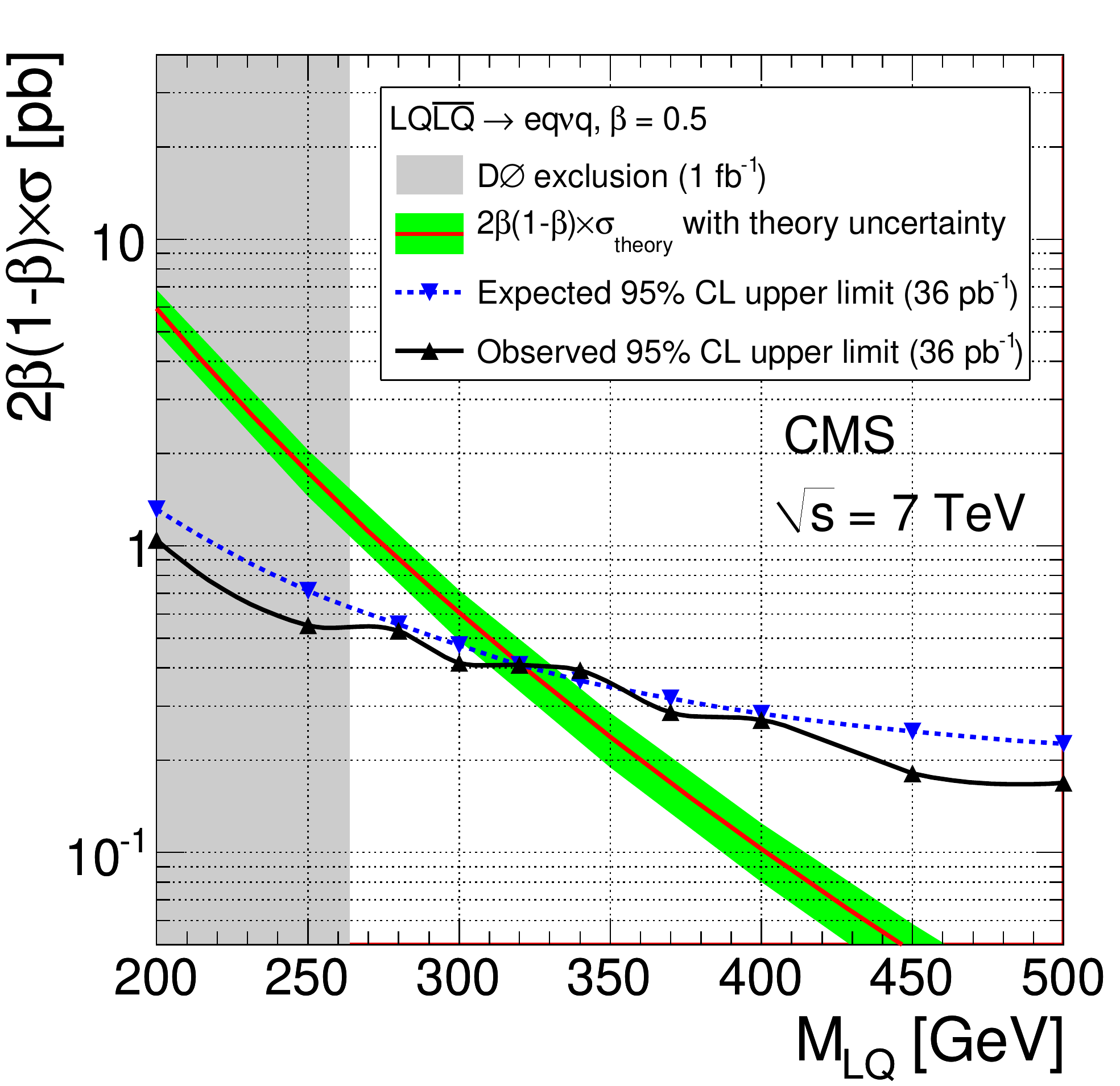}
  \includegraphics[width=0.49\textwidth]{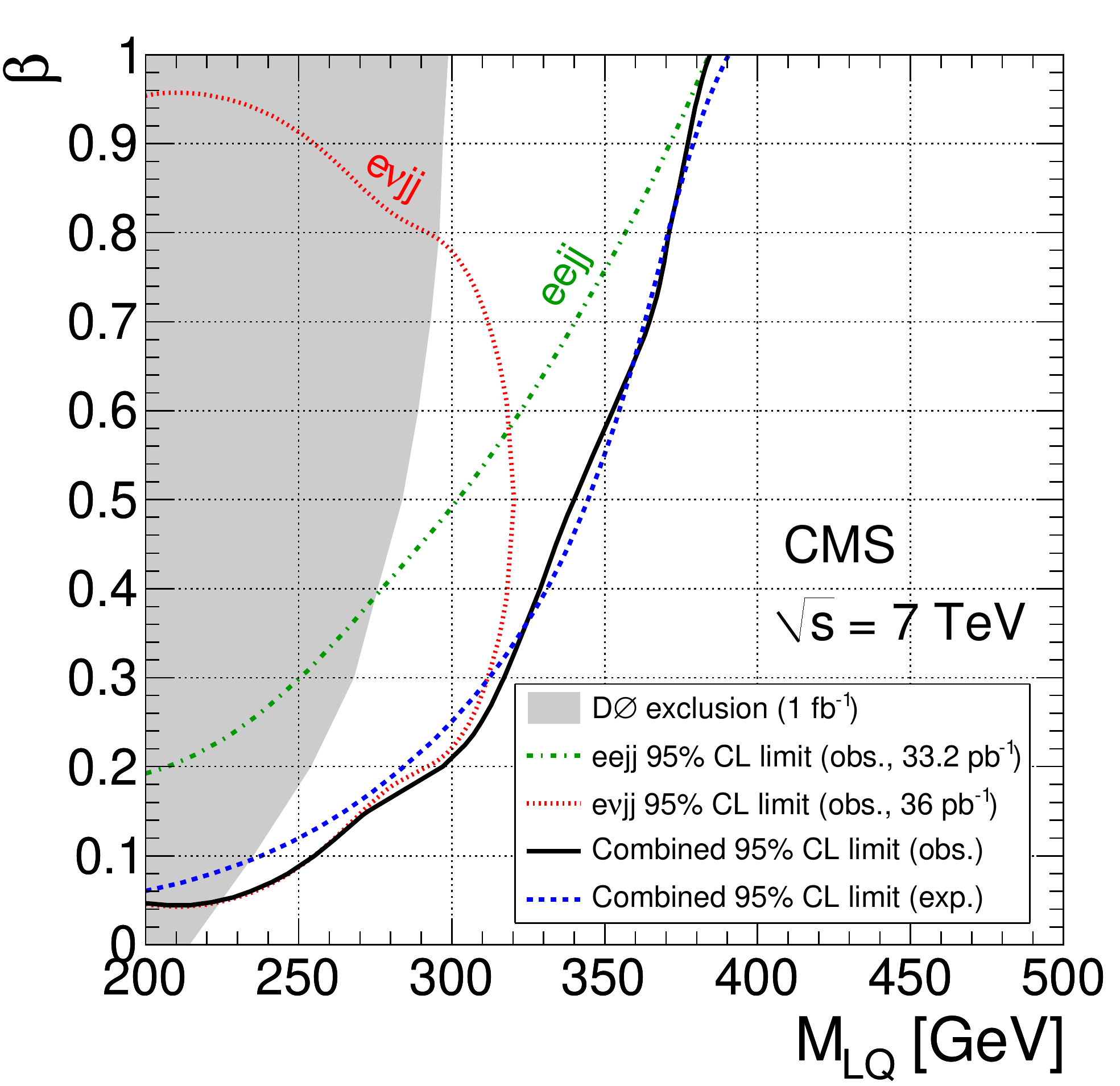}
   \caption{
    (Left)
    The expected and observed upper limits at $95\%$ CL on the LQ pair-production
    cross section times $2\beta(1-\beta)$ as functions of the first generation LQ mass.
    The shaded region is excluded by the current D0~limit for $\beta=0.5$ in the \enujj channel only.
    (Right)
    Observed exclusion limits at 95\% CL on the first generation LQ hypothesis
    in the $\beta$ versus LQ mass plane using the central value of signal cross section, for the
    individual \eejj and \enujj channels, and their combination.
    The combined expected limit is also shown.
    The shaded region is excluded by the current D0~limits, which combine
    results of \eejj, \enujj, and \nunujj decay modes.
    }
  \label{fig:exclusion_xs}
\end{figure*}

Figure~\ref{fig:exclusion_xs} (left)
and Table~\ref{tab:ObsExpUpperLimitOnLQCrossSection} show the $95\%$ confidence level
(CL) upper limit on the LQ pair-production cross section
times $2\beta(1-\beta)$ as a function of the leptoquark mass.
The upper limits are compared to the NLO prediction
of the LQ pair-production cross section~\cite{PhysRevD.71.057503} to
set an exclusion of the first generation scalar LQ mass
smaller than \LQmassLimitBetaHalf, assuming $\beta=0.5$, at the $95\%$ CL.
The theoretical uncertainties on the signal production cross sections
due to the choice of the PDFs
(\PDFUncertaintyOnLQCrossSection~\cite{PhysRevD.71.057503},
calculated using CTEQ6.6~\cite{PhysRevD.78.013004})
and the choice of renormalization and factorization scales
(\theoreticalUncertaintyOnLQCrossSection~\cite{PhysRevD.71.057503},
determined by varying the scales between half and twice the LQ mass)
are shown as a band around the central value in Fig.~\ref{fig:exclusion_xs}
(left).
If the observed upper limit is compared with the lower boundary of this theoretical uncertainty,
the lower limit on the first generation LQ mass for $\beta=0.5$ becomes
\LQmassLimitBetaHalfWithTheoryError.

\begin{table}[htbp]
\caption{Observed and expected $95\%$ confidence level (CL) upper limits on the LQ
pair-production cross section times $2\beta(1-\beta)$ as a function of the leptoquark mass.}
\label{tab:ObsExpUpperLimitOnLQCrossSection}
\begin{center}
\begin{tabular}{|c|c|c|}
\hline
$M_{LQ}$      & \multicolumn{2}{c|}{95\% CL upper limit on $2 \beta (1-\beta) \times \sigma$ {[pb]}} \\
{[GeV]}    & ~~~~~~~~~~Observed~~~~~~~~~~ & ~~~~~~~~~~Expected~~~~~~~~~~ \\
\hline
200      & 1.045 & 1.316 \\
250      & 0.551 & 0.713 \\
280      & 0.528 & 0.553 \\
300      & 0.416 & 0.474 \\
320      & 0.409 & 0.409 \\
340      & 0.392 & 0.364 \\
370      & 0.286 & 0.317 \\
400      & 0.271 & 0.284 \\
450      & 0.181 & 0.248 \\
500      & 0.169 & 0.226 \\
\hline
\end{tabular}
\end{center}
\end{table}

The \enujj channel is combined with the existing CMS results from the
\eejj analysis~\cite{PhysRevLett.106.201802}, 
thereby improving the reach of this search in the
intermediate $\beta$ range.
The likelihoods built for the individual dielectron and electron+neutrino
channels are multiplied.
The same Bayesian approach used to set the individual limits is then applied to the likelihood product
to set the combined limit. While integrating over nuisance parameters, the systematic uncertainties
on signal efficiency and background are
assumed to be fully correlated and the largest
uncertainty amongst the two channels is used.
Figure~\ref{fig:exclusion_xs}
(right) shows the
exclusion limits at 95\% CL on the first generation leptoquark hypothesis in
the $\beta$ versus LQ mass plane, using the central value of the signal cross section, for
the individual dielectron and electron+neutrino channels, and their combination.

%% file: conclusions.tex
\section{Summary}

A search for pair-production of first generation scalar leptoquarks 
in events with an electron, missing transverse energy, 
and at least two jets has been presented.
The contribution of the main backgrounds has been determined by MC studies
and the uncertainty estimated by a comparison with the data.
The number of observed events passing a selection optimized for 
exclusion of the LQ hypothesis is in good agreement with the predictions 
for \sm background processes.
A Bayesian approach that includes treatment of the systematic uncertainties 
as nuisance parameters has been used to set upper limits on the LQ cross section. 
Prior CMS results in the dielectron channel are combined with this 
electron+neutrino search.  A 95\% confidence level combined lower limit is set on 
the mass of a first generation scalar leptoquark at \LQmassLimitBetaHalfComb~GeV 
for $\beta=0.5$. 
These results represent the most stringent direct limits to
date for values of $\beta$ greater than 0.05.

\section*{Acknowledgments}
We wish to thank 
Michael Kr\"amer for providing the 7 TeV NLO LQ pair-production cross sections and
to congratulate our colleagues in the CERN accelerator
departments for the excellent performance of the LHC machine. We thank
the technical and administrative staff at CERN and other CMS
institutes, and acknowledge support from: FMSR (Austria); FNRS and FWO
(Belgium); CNPq, CAPES, FAPERJ, and FAPESP (Brazil); MES (Bulgaria);
CERN; CAS, MoST, and NSFC (China); COLCIENCIAS (Colombia); MSES
(Croatia); RPF (Cyprus); Academy of Sciences and NICPB (Estonia);
Academy of Finland, ME, and HIP (Finland); CEA and CNRS/IN2P3
(France); BMBF, DFG, and HGF (Germany); GSRT (Greece); OTKA and NKTH
(Hungary); DAE and DST (India); IPM (Iran); SFI (Ireland); INFN
(Italy); NRF and WCU (Korea); LAS (Lithuania); CINVESTAV, CONACYT,
SEP, and UASLP-FAI (Mexico); PAEC (Pakistan); SCSR (Poland); FCT
(Portugal); JINR (Armenia, Belarus, Georgia, Ukraine, Uzbekistan); MST
and MAE (Russia); MSTD (Serbia); MICINN and CPAN (Spain); Swiss
Funding Agencies (Switzerland); NSC (Taipei); TUBITAK and TAEK
(Turkey); STFC (United Kingdom); DOE and NSF (USA).

%% file: EXO-10-006-authorlist.tex
\textbf{Yerevan Physics Institute,  Yerevan,  Armenia}\\*[0pt]
S.~Chatrchyan, V.~Khachatryan, A.M.~Sirunyan, A.~Tumasyan
\vskip\cmsinstskip
\textbf{Institut f\"{u}r Hochenergiephysik der OeAW,  Wien,  Austria}\\*[0pt]
W.~Adam, T.~Bergauer, M.~Dragicevic, J.~Er\"{o}, C.~Fabjan, M.~Friedl, R.~Fr\"{u}hwirth, V.M.~Ghete, J.~Hammer\cmsAuthorMark{1}, S.~H\"{a}nsel, M.~Hoch, N.~H\"{o}rmann, J.~Hrubec, M.~Jeitler, W.~Kiesenhofer, M.~Krammer, D.~Liko, I.~Mikulec, M.~Pernicka, H.~Rohringer, R.~Sch\"{o}fbeck, J.~Strauss, A.~Taurok, F.~Teischinger, P.~Wagner, W.~Waltenberger, G.~Walzel, E.~Widl, C.-E.~Wulz
% ...
\vskip\cmsinstskip
% ...
1:~~Also at CERN, European Organization for Nuclear Research, Geneva, Switzerland\\
% ...
